\documentclass[a4paper,10pt]{article}

\pdfoutput=1

\usepackage{cite,amsmath,amsfonts,amsthm,fullpage}
\usepackage{youngtab}

\usepackage{graphics}
\usepackage{miniplot}
\usepackage{subfigure}

\newcommand{\Pf}{\mathop\mathrm{Pf}\nolimits}

\newcommand{\Pa}{\mathop\mathrm{P}\nolimits}

\newcommand{\DP}{\mathop\mathrm{DP}\nolimits}

\newcommand{\ttr}{\texttt{tr}}

\newcommand{\e}{\texttt{e}}

\theoremstyle{plain}

\newtheorem{Lemma}{Lemma}
\newtheorem{Proposition}{Proposition}
\newtheorem{Corollary}{Corollary}
\newtheorem{Remark}{Remark}
\newtheorem{Example}{Example}

\theoremstyle{remark}

\def\tr{\mathrm {Tr}}
\def\ttr{\mathrm {tr}}
\def\det{\mathrm {det}}

\def\bp{\begin{Proposition}}
\def\ep{\end{Proposition}}
\def\bc{\begin{Corollary}}
\def\ec{\end{Corollary}}
\def\bl{\begin{Lemma}}
\def\el{\end{Lemma}}
\def\be{\begin{equation}}
\def\ee{\end{equation}}
\def\br{\begin{Remark}\rm\small}
\def\er{\end{Remark}}
\def\brs{\begin{remarks}.\\ \rm\
\begin{enumerate}}
\def\ers{\end{enumerate}\end{remarks}}
\def\bea{\begin{eqnarray}}
\def\eea{\end{eqnarray}}

\def\bx{\begin{Example}\rm\small}
\def\ex{\end{Example}}

\def\tr{\mathrm {Tr}}
\def\tr{\mathrm {tr}}
\def\det{\mathrm {det}}

\def\res{\mathop{\mathrm {res}}\limits}

\def\&{&{\hskip -20pt}}

\newcount\YDcount\YDcount=0
\def\YDsize{10pt}

\def\YD#1{%
\ifnum#1=0
 \ifnum\YDcount=0 \ifx\varnothing\undefined\emptyset\else\varnothing\fi
 \else\vskip1.4pt\egroup\YDcount=0\fi
\else
 \ifnum\YDcount=0 \YDcount=1\vcenter\bgroup\vskip1pt
 \else\nointerlineskip\fi
 \vbox{\hrule\hbox{\vrule height\YDsize
 \loop\hskip\YDsize\vrule\ifnum\YDcount<#1\advance\YDcount1\repeat}\hrule
 \kern-0.4pt}\expandafter\YD
\fi}

\usepackage[usenames,dvipsnames]{color}
\usepackage{ulem}

\def\pb{\mathbf{p}}
\def\tb{\mathbf{t}}

\def\xb{\mathbf{x}}

\begin{document}

\author{E. N. Antonov\thanks{Petersburg Nuclear Physics Institute named by B.P.Konstantinov of NRC ``Kurchatov Institute'',  1 mkr. Orlova Roshcha, Gatchina, Leningradskaya Obl, 188300, Russia,
email: antonov@thd.pnpi.spb.ru } \\A. Yu. Orlov\thanks
{
Institute of Transmition Information Problems RAS, Moscow 127051 Russia, Bolshoy Karetny per. 19, build. 1;
Institute of Oceanology, Nahimovskii Prospekt 36,
Moscow 117997, Russia, email: orlovs@ocean.ru
}
}
\title{New solvable two-matrix model and BKP tau function}

\date{}
\maketitle

\begin{abstract}
We present exactly solvable modifications of the two-matrix Zinn-Justin-Zuber model and write it as a tau function. The grand partition function of these matrix integrals  is written as the fermion expectation value. The perturbation theory series is written out explicitly in terms of series in strict partitions. The related string equations are presented.
\end{abstract}

\bigskip
\medskip


\section{Introduction \label{Introduction}}

This note was initiated by the work on the generalized Kontsevich model \cite{MMq} and further discussions with the authors. The perturbation theory series for 
the partition function of this model was written in a very compact form as a sum over strict partitions of 
a pair of projective Schur functions. It was followed by the work \cite{Alex} where the similar series appeared
for the different model (BGW model). The sources of interest to series in projective Schur functions can be
found in 
\cite{Lee2018},\cite{Serbian},\cite{MMN2019},\cite {Alex},\cite{MMNO},\cite{Alex2},\cite{Alex2},\cite{MM-genQ},\cite{MMZ},\cite{MMZ-2},
\cite{MMMZ}. There are some earlier works on the series, 
see \cite{TW},\cite{O-hyp-BKP},\cite{ONimmo},\cite{Sho},\cite{LO-2008},\cite{HLO}. On the projective Schur 
functions and the representation theory of the supersymmetric group 
$q(N)$ and the symmetric group $S_n$,
see \cite{Serg},\cite{Ivanov},\cite{EOP},\cite{Strembridge}. 
 The appearance of these functions in the integrable
models was presented in \cite{You} and \cite{Nim}.

If a matrix integral is a tau function as a function of its coupling constants, we call it solvable. 
As far as we know, the first solvable (in this sense) matrix model was presented in the preprint of \cite{GMMMO}; see other examples in \cite{KMMOZ}, \cite{GKM} and \cite{KMMM}. Then we should point  out the work \cite{L1}.

Here, we present and compare two families of solvable matrix integrals.
The second family is completely new and is related to the KP hierarchy on the root system B (the BKP hierarchy,
which was introduced in \cite{DJKM1}).

\section{Models of two unitary matrices}

\subsection{Standard and modified models of two unitary matrices}

In \cite{ZinZub} the following integral over two unitary matrices was studied
\be\label{I1}
I_1(\tb,\tb^*)=C\int_{\mathbb{U}_N\times \mathbb{U}_N} e^{c\,\tr \left(U_1^{-1}U_2^{-1}\right)+\sum_{n>0}
\frac 1n \left( t_n \tr U_1^n  + t^*_n \tr U_2^n   \right)}d_*U_1 dU_2^*
\ee
where $d_*U$ is the Haar measure of the unitary group $\mathbb{U}_N$. The number $c$ and the sets
$\tb=(t_1,t_2,\dots),\,\tb^*=(t_1^*,t_2^*,\dots)$ play the role of coupling constants in the model,
and $C$ is a normalization constant: $CI(0,0)=1$.
It is shown that $I_1$ can be written explicitly as a series of the Schur polynomials 
as functions in the coupling constants 
over partitions as follows:
\be
I_N(\tb,\tb^*) =\sum_\lambda s_\lambda(\tb)s_\lambda(\tb^*)\prod_{(i,j)\in\lambda} \frac{c}{(N-i+j)!}
\ee
where $(i,j)$ are the coordinates of the node of the Young diagram $\lambda$. The sum ranges over
all Young diagrams of height do not accede $N$, that is, $j=1,2,\dots$ and $i\le N$.
It can be shown that this series is an example of the KP tau functions, also known as hypergeometric tau functions \cite{KMMM}, \cite{OS-TMP}, which admit relatively simple determinant forms.

\paragraph{Modified family}.
In \cite{O-2004-New} (see also Appendix A in \cite{HO-2005}), the following generalization of
the integral (\ref{I1}) was introduced, namely, one can make a replacement for the term responsible
for the interaction between matrices $U_1$ and $U_2$:
\be\label{replacement1}
e^{c\,\tr U^{-1}_1U^{-1}_2}\,\to \, \tau\left(N; c\,U^{-1}_1U^{-1}_2; f\right)
\ee
where $\tau\left(c\,U^{-1}_1U^{-1}_2; f\right)$ is defined by the choice of the function $f$
as follows:
\be\label{tau-hyp}
\tau\left(N;  c\,U^{-1}_1U^{-1}_2;f\right):=\sum_{\lambda} \prod_{i<j\le N}(\lambda_i-\lambda_j-i+j)
s_\lambda\left(c\,U^{-1}_1U^{-1}_2\right)\prod_{i\le N} 
f(\lambda_i-i+N)
\ee
Examples:
\be
{\rm if} \,\,f(x)=\frac {1}{\Gamma(x+1)}\,\,{\rm then}\,\,\tau\left(N;  c\,U^{-1}_1U^{-1}_2;f\right)=e^{c\,\tr \left(U_1^{-1}U_2^{-1}\right)}
\ee
\be
{\rm if} \,\,f(x)=\frac {\Gamma(x+a)}{\Gamma(a)\Gamma(x+1)}\,\,{\rm then}\,\,\tau\left(N;  c\,U^{-1}_1U^{-1}_2;f\right)=\det\left(1-c\,U_1^{-1}U_2^{-1}\right)^{-a}
\ee

The wonderful property of such tau function is the following determinantal representation:
\be
\int_{\mathbb{U}_N}\tau\left(c\,UU^{-1}_1U^{-1}U^{-1}_2;f\right)d_*U
=
\frac{\det[\tau(1; c \,u^{-1}_iv^{-1}_j;f)]_{i,j\le N}}{\prod_{i<j\le N}(u^{-1}_i-u^{-1}_j)(v^{-1}_i-v^{-1}_j)}
\ee
where
\be
\tau(1; c \,u^{-1}v^{-1};f)]=1+f(1)u^{-1}v^{-1}+f(2)u^{-2}v^{-2}+\cdots
\ee
which allows us writing down
the perturbation series for the generalized model in the form of another matrix integral:
\be\label{I1-f-series}
I_N(\tb,\tb^*;f)=
C\int_{\mathbb{U}_N\times \mathbb{U}_N}\tau\left(N; c \,U^{-1}_1U^{-1}_2;f\right)
e^{\sum_{n>0}
\frac 1n \left( t_n \tr U_1^n  + t^*_n \tr U_2^n   \right)}d_*U_1 dU_2^*
\ee
\be
=\sum_{\lambda\atop \ell(\lambda)\le N} s_\lambda(\tb)s_\lambda(\tb^*)
\prod_{i} f(\lambda_i-i+N) c^{\lambda_i}
\ee
which is the Toda lattice \cite{Mikhailov} tau function \cite{UT},\cite{Takasaki-Schur},\cite{Takasaki2018}
as presented in \cite{KMMM} and carefully studied 
in \cite{OS-2000}. In \cite{HO-2005}, the integral (\ref{I1-f-series}) was written as the fermionic
vacuum expectation value. See Appendix \ref{I(f)}, where
the linear equations (sometimes known as string ones) for (\ref{I1}) and (\ref{I1-f-series}) are written down.

\section{New models of two unitary matrices}

 Let us consider a modification of the models mentioned above.
Let us consider the following model: 
\be\label{J1}
J_N(\pb^{(1)},\pb^{(2)})=C_N\int_{\mathbb{U}_N\times \mathbb{U}_N} e^{c\,\tr \left(U_1^{-2}U_2^{-2}\right)+\sum_{n=1,3,5,\dots}
\frac 2n \left( p^{(1)}_n \tr U_1^n  + p^{(2)}_n \tr U_2^n   \right)} 
d\mu_1(U_1) d\mu_2(U_2)
\ee
where
\be
d\mu_i(U)=\det\left(U^{-\frac12(N^2-N)+\kappa_i}  \right)d_*U_i,\quad i=1,2
\ee
For the sake of simplicity,
we choose $\kappa_1=\kappa_2=0$ (all calculations below can be reproduced also in case $\kappa_i \neq 0$,
but all formulas take up noticeably more space in this case).
Here, the constant $C_N$ is chosen to ensure the normalization $J_N(0,0)=1$.
In this model, the parameters $\pb^{(1)}=(p^{(1)}_1,p^{(1)}_3,\dots),\,\pb^{(2)}=(p^{(2)}_1,
p^{(2)}_3,\dots)$ play the role of coupling constants.

To study (\ref{J1}) we apply the Harish-Chandra-Itzykson-Zuber relation:
\be\label{HCIZ}
\int_{\mathbb{U}_N} e^{\tr \left(UAU^{-1}B\right)} d_* U = \frac{\det \left[e^{a_kb_l}\right]_{1\le k,l\le N} }{\prod_{k<l}(a_k-a_l)(b_k-b_l)}
\ee
where $a_i$ and $b_i$, $i=1,\dots , N$ are the eigenvalues of matrices $A$ and $B$.
Then we rewrite $J_N$ as the integral in eigenvalues, also using the symmetry property of the 
integrand:
\be\label{J1eig}
J_N(\pb^{(1)},\pb^{(2)})=\tilde{C}_N\oint \dots \oint \prod_{i<j}\frac{(u_i-u_j)(v_i-v_j)}{(u_i+u_j)(v_i+v_j)} \prod_{i=1}^N e^{c\,\tr \left(u_i^{-2}v_i^{-2}\right)+\sum_{n=1,3,5,\dots}
\frac 2n \left( p^{(1)}_n u_i^n  + p^{(2)}_n v_i^n   \right)}  \frac{du_i}{u_i} \frac{dv_i}{v_i}
\ee
To obtain this form, we use (\ref{HCIZ}) as well as the well-known explicit form of the expression
for the Haar measure $\mathbb{U}_N$, which contains the factors $d_*U_1 \sim \prod_{i<j} |u_i-u_j|^2\prod_{k}\frac{du_k}{u_k}$ and
$d_*U_2\sim \prod_{i<j}|v_i-v_j|^2 \prod_{k} \frac{dv_k}{v_k}$.

For the next step, we use the following analogue of the Cauchy-Littlewood relation \cite{Mac}:
\be\label{Cauchy}
e^{\sum_{n=1,3,5,\dots} \frac 2n p_n\tr \left(X^n\right)} = \sum_{\alpha\in DP\atop \ell(\alpha)\le N} Q_\alpha(\pb)
Q_\alpha\left(\pb(X)\right)
\prod_{i=1}^{\ell(\alpha)} 2^{-\alpha_i}
\ee
where $X$ is a matrix and 
where $Q_\alpha$ is the so-called projective Schur function, see Appendix. This is a polynomial
in the variables $p_1,p_3,\dots$ and $$\alpha=(\alpha_1,\alpha_2,\dots,\alpha_k),\,\alpha_1>\cdots >\alpha_k\ge 0,
\quad k=1,2,\dots$$
is the multi index of this 
multivariable polynomial. Such sets are called strict partitions, and the set of all strict partitions
(or the same: the set of Young diagrams with distinct lengths of rows $\alpha_1>\alpha_2>\cdots$)
we denote $DP$ as in \cite{Mac}. The notation
$Q_\alpha\left(\pb(X) \right)$ means that here the arguments $p_1,p_3,\dots$ are not free parameters but  chosen to be equal
Newton sums of the eigenvalues of the matrix $X$:
\be
p_n=p_n(X):=\tr \left(X^n\right),\quad n \,{\rm is\, odd}
\ee
Thus, the polynomial $Q_\alpha\left(\pb(X) \right)$ is a symmetric function in the eigenvalues of $X$.
For the sake of simplicity, we shall write $Q_\alpha(X)$ instead of
$Q_\alpha\left(\pb(X) \right)$ having in mind that a capital letter serves for a matrix.

At last, one can prove (see Appendix \ref{scalar}) that
\be\label{QQproduct}
\frac{1}{\left(2\pi i \right)^{2N}}\oint \dots \oint  Q_\alpha(U_1)Q_\beta(U_2)
\prod_{i<j}\frac{(u_i-u_j)(v_i-v_j)}{(u_i+u_j)(v_i+v_j)} \prod_{i=1}^N 2^{c u_i^{-2}v_i^{-2}}
\frac{du_i}{u_i} \frac{dv_i}{v_i}
\ee  
$$
=\begin{cases}
\delta_{\alpha,\beta} 2^{2\ell(\alpha)}\prod_{i=1}^{\ell(\alpha)}\frac{c^{\alpha_i}}{(\frac12 \alpha_i)!},\,
{\rm if\,each}\,\alpha_i\,
{\rm is\, even}\\
0\quad {\rm otherwise} 
\end{cases}
$$
see Appendix. From (\ref{J1eig}),(\ref{Cauchy}) and (\ref{J1QQ}) we obtain
\be\label{J1QQ}
J_N(\pb^{(1)},\pb^{(2)})=\sum_{\alpha\in DP\atop \ell(\alpha)\le N} Q_{2\alpha}(\pb^{(1)})
Q_{2\alpha}(\pb^{(2)})
\prod_{i=1}^{\ell(\alpha)} c^{2\alpha_i} f(\alpha_i),
\ee
where
\be\label{f1}
f(\alpha_i)=\frac{1}{\alpha_i!}
\ee
compare to (\ref{I1}). The series in the right-hand side can be obtained by a limiting procedure
from the hypergeometric BKP tau function studied in \cite{O-hyp-BKP}.

If we generalize integral (\ref{J1}) with the help of replacement (\ref{replacement1}),
namely
\be\label{J1-f}
J_N(\pb^{(1)},\pb^{(2)};f)=C\int_{\mathbb{U}_N\times \mathbb{U}_N} \tau(N;cU_1^{-2}U^{-2}_2;f)e^{\sum_{n=1,3,5,\dots}
\frac 2n \left( p_n^{(1)} \tr U_1^n  + p^{(2)}_n \tr U_2^n   \right)} 
d\mu_1(U_1) d\mu_2(U_2)
\ee
we get (\ref{J1QQ}).

\paragraph{Fermionic form for the grand partition function}

The grand partition function for the sum (\ref{J1QQ}) is
\be\label{grandJ1}
J(\pb^{(1)},\pb^{(2)},\zeta)=\sum_N J_N(\pb^{(1)},\pb^{(2)})\zeta^N=\sum_N \zeta^N
\sum_{\alpha\in DP\atop \ell(\alpha)\le N} Q_{2\alpha}(\pb^{(1)})Q_{2\alpha}(\pb^{(2)})
\prod_{i=1}^{\ell(\alpha)} \frac{c^{2\alpha_i}}{\alpha_i!},
\ee
where we insert (\ref{J1QQ}),(\ref{f1})
for the model (\ref{J1}).

This series can be written as fermionic vacuum expectation value:
\be\label{J1vev}
J(\pb^{(1)},\pb^{(2)},\zeta)=\langle 0| \mathbb{T}\e^{\zeta F(\pb^{(1)},\pb^{(2)})} |0 \rangle,
\quad
F(\pb^{(1)},\pb^{(2)})= -\frac{1}{4\pi^2}\oint\oint e^{cu^{-2}v^{-2}}\phi^{(1)}(u,\pb^{(1)})\phi^{(2)}(v,\pb^{(2)})\frac{du dv}{uv} 
\ee
 where
\be
\phi^{(i)}(\pb,z)=e^{\sum_{n=1,3,5,\dots} \frac 1n p_n z^n}\phi^{(i)}(z),\quad i=1,2,
\ee
are neutral fermions:
\be\label{canonical-fields}
[\phi^{(i)}(z_1),\phi^{(i)}(z_2)]_+=\sum_{n\,{\rm odd}} (-1)^n \frac{z_1^n}{z_2^n},\quad 
[\phi^{(1)}(z_1),\phi^{(2)}(z_2)]_+=0
\ee
(Fermi fields $\phi^{(i)}(z,\pb)$ obey the same anticommutation relations.)
Here $\mathbb{T}$ is the chronological ordering of fermions in 2D Eucledian space in the Taylor series
of the exponential (which means
the ordering of the absolute values of arguments $|z_a|>|z_b|$ if $\phi^{(i)}(z_a)$ 
is placed to the left of $\phi^{(i)}(z_a)$ and the sign factor is taken into account). In further,
we shall omit the symbol $\mathbb{T}$ keeping it in mind as the well-known standard point.
This expression is rather similar to the vacuum expectation value written down in \cite{AntOr} for the  Fatteev-Frolov-Schwartz instanton sum in $\sigma$-model.
Fermi modes defined by $\phi^{(i)}(z)=\sum_{j\in\mathbb{Z}} \phi^{(i)}_j z^j$ obey 
\be\label{canonical}
[\phi^{(i_1)}_{j_a},\phi^{(i_2)}_{j_b}]_+=(-1)^{j_a}\delta_{i_1,i_2}\delta_{j_a+j_b}
\ee
We notice that $F(0,0)=\sum_{i\in \mathbb{Z}} \frac{\phi^{(1)}_{2i}\phi^{(2)}_{2i}}{i!}$.
Now, attention, we use the approach of Kac-van de Leur \cite{KvdLbispec} where the vacuum vectors 
are chosen as follows:
\be
\phi^{(i)}_0|0\rangle=\frac{1}{\sqrt{2}}|0\rangle,\quad
\langle 0|\phi^{(i)}_0 =\frac{1}{\sqrt{2}}\langle 0|,\quad i=1,2
\ee
\be
\phi_n|0\rangle  = 0,\quad \langle 0|\phi_{-n}=0,\quad n<0
\ee
This means that
\be
\langle 0|\phi(z,\pb)|0\rangle= \frac{1}{\sqrt{2}}e^{\sum_{n=1,3,\dots} \frac2n p_n z^n}
\ee
and for $N=1,2,3,\dots$ we get 
\be\label{corr-p-dependent'}
\langle 0| \phi^{(i)}(z_1,\pb)\cdots\phi^{(i)}(z_N,\pb) |0\rangle =
2^{\frac N2}
\prod_{i<j\le N
}\frac{z_i-z_j}{z_i+z_j}
 e^{\sum_{m=1,3,\dots}\sum_{i=1}^N
\frac 2m p_m z_i^m  }
\ee
\be\label{corr-p-dependent}
=
2^{\frac N2}\prod_{i<j\le N}\frac{z_i-z_j}{z_i+z_j} \sum_{\alpha\in DP\atop \ell(\alpha)\le N} 2^{-\ell(\alpha)} Q_\alpha(\pb)Q_\alpha(Z)
\ee
where $z_1,\dots,z_N$ are eigenvalues of an $N\times N$ matrix $Z$.
The first factor in (\ref{corr-p-dependent}) we got by the Wick theorem,
and the second factor  in (\ref{corr-p-dependent}) was obtained from (\ref{Cauchy}). However,
it goes back to 
the work \cite{You}, which, in turn, was based on \cite{DJKM1}. The vacuum expectation value (\ref{J1vev})
is an example of the two-component BKP tau function introduced in \cite{DJKM1},\cite{KvdLbispec}. The by-product 
of this statement is the following bilinear  equations  (the same: Hirota equations) for the integral (\ref{J1}). The Hirota equations for the two-component BKP tau function can be found in
\cite{KvdLbispec}. The grand partition function (\ref{J1vev}) solves these bilinear equations as
well as any other BKP tau function. The special tau function is selected by constraints written in the form
of linear differential equations, see below the paragraph on  string equations.

Neutral Fermi fields $\phi^{i}(z)$ can be realized as anticommuting operators $V(z,{\hat\pb}^{(i)})$ acting in the bosonic Fock
space $\cal{F}$ formed by polynomials in the variables $\pb^{(1)},\pb^{(2)}$ multiplied by 
$\frac{1}{2}(1+\eta_1)(1+\eta_2)$ where $\eta_i,\,i=1,2$ are odd Grassmannian numbers: $\eta_i^2=0$.
The (right) vacuum vector is $\frac{1}{2}(1+\eta_1)(1+\eta_2)$,
creation operators are $p^{(1)}_n$ and $p^{(2)}_n$, $n$ odd,
and annihilation operators are equal to the derivatives $n\partial_{p^{(1)}_n}$ and $n\partial_{p^{(2)}_n}$.

Let 
\be\label{vertex-B}
V(z,\hat{\pb}^{(i)})=\frac{\eta_i+\frac{\partial}{\partial \eta_i}}{\sqrt{2}}
e^{\sum_{m\in\mathbb{Z}^+_{odd}} \tfrac2m z^m p^{(i)}_m}e^{-\sum_{m\in\mathbb{Z}^+_{odd}}  z^{-m}
\tfrac {\partial}{\partial p^{(i)}_m}},\quad |z|=1,
\ee
be the vertex operator as it was introduced in \cite{KvdLbispec}.
The symbol $\hat{\pb}^{(i)}$ denotes the set of two collections: 
$\eta_i,p^{(i)}_1,p^{(i)}_3,p^{(i)}_5,\dots$
and $\frac{\partial}{\partial \eta_i},\frac {\partial}{\partial p^{(i)}_1},\frac {\partial}{\partial p^{(i)}_3},\dots$.

One can verify that vertex operators $V^{(i)}(z)$ satisfy relations (\ref{canonical-fields}). 

The bosonization formulae result it the following representation for $J(\pb,\pb^*,\zeta)$:
\be\label{bosonicJ1}
J(\pb^{(1)},\pb^{(2)},\zeta)=g^{\rm Bos}({\hat\pb}^{(1)},{\hat\pb}^{(2)},\zeta) \cdot 1,
\ee
and
\be
g^{\rm Bos}({\hat\pb}^{(1)},{\hat\pb}^{(2)},\zeta) = 
e^{\zeta\oint\oint e^{cu^{-2}v^{-2}}V(u,{\hat\pb}^{(1)})V(v,{\hat\pb}^{(2)})\frac{du dv}{uv} }
\ee

\paragraph{String equations}.
The general construction of the algebra $W^B_{1+\infty}$ is presented in \cite{Leur1994},\cite{Leur1996}. 

Following the standard procedure, one can expand the product of the two vertex operators in the generators of
$W^B_{1+\infty}$ algebra:
\be
\frac 12 V(ze^{\tfrac y2},\hat\pb)   V(-ze^{-\tfrac y2},\hat\pb) - \frac 12 \frac{1+e^{-y}}{1-e^{-y}} =
\frac 14 \frac{e^y+1}{e^y-1}\left(\vdots e^{\theta(ze^{\tfrac y2})+\theta(-ze^{-\tfrac y2})} \vdots -1\right)=
\ee
\be
=\frac 14 \frac{e^y+1}{e^y-1}
\sum_{k>0}\frac{1}{k!}\vdots \left(
 \sum_{m\in\mathbb{Z}_{odd}}\theta_m(\hat\pb) z^m \left(e^{\tfrac {my}{2}} -e^{-\tfrac {my}{2}}\right)
\right)^k \vdots
=:\sum_{m\in\mathbb{Z},n\ge 0} \frac {1}{n!} y^{n} z^m \Omega_{\textsf{B}}(m,n,\hat\pb)
\label{parted-vertices}
\ee
where $\vdots a\vdots$ means the bosonic normal ordering (which means that all derivatives
$\partial_{p_i}$ are moved to the right of $p_i$) and where
\be
2\theta(z,\hat\pb) := \sum_{m\in\mathbb{Z}^+_{odd}} \tfrac 2m  z^m p_m
- \sum_{m\in\mathbb{Z}^+_{odd}}   z^{-m} \tfrac {\partial}{\partial p_m}
\ee

Equivalently, one can write
\be
 \Omega_{\textsf{B}}(m,n,\hat\pb)  = -\res_z V(-z,\hat\pb) z^{-\tfrac m2}\left(D^n z^{-\tfrac m2} V(z,\hat\pb)\right)
\frac{dz}{z},\quad D=z\frac{\partial}{\partial z}
\ee

As follows from the left-hand side of this formula, $\Omega_{mn}$ vanishes when $n$ and $m$ have
the same parity.
 In particular, we have
$$
\Omega_{\textsf{B}}(0,1,\hat\pb)=\sum_{n>0} n p_n\partial_n
$$
\be
\Omega_{\textsf{B}}(0,3,\hat\pb) = \frac 12 \sum_{n>0} n^3 p_n\partial_n +
\frac {1}{2}\sum_{n>0} n p_n\partial_n
+4\sum_{n_1,n_2,n_3\,{\rm odd}} p_{n_1}p_{n_2}p_{n_3}  (n_1+n_2+n_3)\partial_{n_1+n_2+n_3} +
\ee
\be
+3\sum_{n_1+n_2=n_3+n_4\,{\rm odd}} p_{n_1}p_{n_2} n_3n_4\partial_{n_3}\partial_{n_4} +
\sum_{n_1,n_2,n_3\,{\rm odd}} p_{n_1+n_2+n_3}\partial_{n_1}\partial_{n_2}\partial_{n_3}
\ee

The fermionic counterpart of (\ref{parted-vertices}) is much simpler:
\be\label{::}
 \frac 12 :\phi(ze^{\tfrac y2})\phi(-ze^{-\tfrac y2}): = \frac 12
\sum_{m,j\in\mathbb{Z}} z^{m} e^{\tfrac y2(m+2j)}(-1)^j:\phi_{m+j}\phi_{-j}: =
\sum_{m\in\mathbb{Z}\atop n\ge 0} \frac{1}{n!}  y^n z^m \Omega_\textsc{f}(m,n)
\ee
where $:a:$ serves for the fermionic normal ordering (namely, in formula (\ref{::}) one can 
replace each $:a:$ by $a-\langle 0|a|0\rangle$).
Again, as follows from the left-hand side of this formula, $\Omega_{mn}=0$ when $n$ and $m$ have the same parity.
One gets
\be\label{Omega}
\Omega_\textsc{f}(m,n)=\frac 12 \sum_{j\in\mathbb{Z}}  \left(\tfrac m2+j\right)^n(-1)^j:\phi_{m+j}\phi_{-j}:
\ee
\be
=\res_z \left(  z^{-\frac m2}\cdot D^n\cdot z^{-\frac m2} 
\cdot \phi(z)\right) \phi(-z)\frac {dz}{z},\quad D=z\frac{\partial}{\partial z}.
\ee
Among operators (\ref{Omega}), we have the Virasoro ones, whose fermionic form is
\be
L^{\textsc{f}}_{m}:=\Omega_\textsc{f}(-2m,1)=
 \frac 12 \sum_{j\in\mathbb{Z}} \left( j-m\right)
(-1)^j \phi_{j-2m}\phi_{-j} 
\ee
and operators
\be
\Omega_\textsc{f}(0,n)=
\frac 12 \sum_{j\in\mathbb{Z}}(-1)^j j^n:\phi_{j}\phi_{-j}:=
\sum_{j=1,3,\dots}(-1)^j j^n\phi_{j}\phi_{-j},\quad n\,{\rm odd}.
\ee
It is known that
$$
\Omega_\textsc{b}(0,n,\hat\pb)Q_\alpha(\pb)=e_\alpha Q_\alpha(\pb),\quad
e_\alpha =\sum_i \alpha_i^n,\quad n=1,3,5,\dots
$$
and therefore we get the following string equation:
\be\label{string_diag}
\left(\Omega_\textsc{b}(0,n,\hat\pb^{(1)})-\Omega_\textsc{b}(0,n,\hat\pb^{(2)})\right)
J_N(\pb^{(1)},\pb^{(2)})=0,\quad n=1,3,5,\dots
\ee
Each of the equations (\ref{string_diag}) characterizes the series (\ref{J1QQ}), where both of the projective Schur functions in the products of pairs
are labeled with the same strict partition, but in no way does it characterize the prefactor 
$\prod_i\frac{2^{-2\alpha_i} c^{2\alpha_1}}{ \alpha_i!}$ in the right-hand side of (\ref{J1QQ}).
To do it, we will write down constraints for the grand partition function, which can be obtained
by the usage of different $BW_{1+\infty}$ elements. We describe this procedure in short.

First, for each given integer $m$, we introduce a special combination of $\Omega_{\textsc{f}}(2m,n),\,n=1,3,\dots$ as follows:
\be\label{M}
{\cal{M}}^{(i)}_{\textsc{f}}(2m,G) =\frac 12 \sum_{j\in\mathbb{Z}} y_{m,G}\left(j+ m \right) 
(-1)^j:\phi^{(i)}_{2m+j}\phi^{(i)}_{-j}:
\ee
\be
=\res_z :\left(  z^{-m}\cdot y_{m,G}(D)\cdot z^{-
m} 
\cdot \phi^{(i)}(z)\right) \phi^{(i)}(-z):\frac {dz}{z}
\ee
where we define $y_{m,G}$ as the following polynomial function of odd degree:
\be\label{y}
y_{m,G}(x)=x\left(x^2-m^2 \right)\left(x^2-(m+1)^2 \right)\cdots 
\left(x^2-(2m-1)^2 \right)G\left(x^2\right),\quad x=m+j
\ee
where  $G(x^2)$ is an arbitrary-chosen polynomial of $x^2$.
This combination of $BW_{1+\infty}$ elements is chosen to provide the property
\be\label{ann-vacuum}
{\cal{M}}^{(i)}_{\textsc{f}}(2m,G)|0\rangle =0,\quad i=1,2
\ee
for any integer $m$ and for any polynomial $G(x^2)$. Indeed,
if $m$ is negative, then each of $\phi_{j+2m}\phi_{-j}$ from the sum in the right-hand side of (\ref{M}) annihilates $|0\rangle$. If $m > 0$, the situation is as follows:
we have
\[
 \frac12\sum_{j\in\mathbb{Z}}  
(-1)^j\phi^{(i)}_{2m+j}\phi^{(i)}_{-j} |0\rangle = \phi^{(i)}_{2m}\phi^{(i)}_{0}|0\rangle +
\dots +(-1)^{m-1} \phi^{(i)}_{m+1}\phi^{(i)}_{m-1}|0\rangle \neq 0,
\]
however
 ${\cal M}^{(i)}_{\textsf{f}}(2m,G)$ annihilates the vacuum vector due to the vanishing of 
 $y_{m,G}(m+i)$ at points $i=0,i=1,\dots, i=m-1$, see (\ref{y}).
Next, by (\ref{canonical}) one can also verify that for $m\ge 0$ we obtain
\be\label{commuting}
 [{\cal{M}}^{(1)}_{\textsc{f}}(2m,G) - {\cal{M}}^{(2)}_{\textsc{f}}(-2m,\tilde{G}) ,
 F(0,0)]=0
 \ee
 where $\tilde{G}(x^2)$ is also a polynomial in $x^2$ and defined as follows:
 \be\label{c}
 \tilde{G}(x^2)=G(x^2)c_{\rm even}(x),\quad c(x)=\frac{\Gamma(x+1+ m)}{\Gamma(x+1)}=c_{\rm even}(x)+c_{\rm odd}(x)
 \ee
 where polynomial $c_{\rm even}(x)=c_{\rm even}(-x)$ $c_{\rm odd}(x)=-c_{\rm odd}(-x)$.
 Say, for $m=1$ we have $c(x)=x+1$, thus $c_{\rm even}=1$ and $y_{1,G}(x)=y_{-1,\tilde{G}}(x)=x(x^2-1)G$.

 Clearly, thanks to the symmetry between $\phi^{(1)}$ and $\phi^{(2)}$ in (\ref{grandJ1}) one can also write
 $$
 [{\cal{M}}^{(2)}_{\textsc{f}}(2m,G) - {\cal{M}}^{(1)}_{\textsc{f}}(-2m,\tilde{G}) ,
 F(0,0)]=0
 $$
 
 The properties (\ref{ann-vacuum})-(\ref{commuting}) and the bosonization procedure
 result in (string) equations on the grand partition function:
 \be\label{string} 
 \left({\cal{M}}_{\textsc{b}}(-2m,{\tilde{G}},\hat\pb^{(2)}) - 
 {\cal{M}}_{\textsc{b}}(2m,G,\hat\pb^{(1)}) \right)J(\pb^{(1)},\pb^{(2)},\zeta)=0
 \ee
 with
 \be\label{M,m}
 {\cal{M}}_{\textsc{b}}(2m,G,\hat\pb^{(1,2)})=
 \res_z \left(  z^{-m}\cdot y_{m,G}(D)\cdot z^{-m} 
\cdot V(z,\hat\pb^{(1,2)})\right) V(-z,\hat\pb^{(1,2)})\frac {dz}{z}
 \ee 
 \be\label{M,-m}
 {\cal{M}}_{-\textsc{b}}(-2m,Gc_{\rm even},\hat\pb^{(2,1)})=
 \res_z \left(  z^{m}\cdot y_{-m,Gc_{\rm even}}(D)\cdot z^{m} 
\cdot V(z,\hat\pb^{(2,1)})\right) V(-z,\hat\pb^{(2,1)})\frac {dz}{z}
 \ee 

\paragraph{On the Pfaffian form}. Let us write $\pb$ instead of $\pb^{(1)}$. Let us choose
 $p^{(2)}_n=p_n(y):=\sum_{i=1}^{2k} y_i^n,\,n=1,3,\dots$. For this choice one can apply the bosonization formula and the 
Wick theorem and get the following pfaffian representation:
\be\label{pf1}
 J(\pb,\pb(y),\zeta)=\prod_{i<j}^{2k}\left(\frac{y_i+y_j}{y_i-y_j} \right)
 \Pf \left[J\left(\pb^{(1)},\pb(y_i,y_j),\zeta\right)\right]_{i,j}
\ee
where 
\be
J\left(\pb,\pb(y_i,y_j),\zeta\right)=\sum_N \int_{U_1\times U_1}
e^{
c\,\tr \left(U_1^{-2}U_2^{-2}\right)+\sum_{n=1,3,\dots}
\frac 2n \left( p^{(1)}_n \tr U_1^n\right)}\det\frac{(1-y_iU_2)(1-y_jU_2)}{(1+y_iU_2)(1+y_jU_2)}  d_*U_1 dU_2^*
\ee
And by the same method, one can write down the answers as the Pfaffian of a block $2k+2l\times 2k+2l$
matrix in case $p^{(1)}_n=\sum_{i=1}^{2k} y_i^n,\,p{(2)}_n=\sum_{i=1}^{2l} z_i^n\,n=1,3,\dots$. We omit
this specious expression and leave it as quite a tedious exercise for the reader, see \cite{paper3} where such calculations are made.

\section{Cauchy-type interaction}

Next, we consider the integral
\be\label{J2}
K_N(\pb^{(1)},\pb^{(2)})=C\int_{\mathbb{U}_N\times \mathbb{U}_N} 
\det\left(1-c U_1^{-2}U_2^{-2}\right)^{-a}
e^{\sum_{n=1,3,5,\dots}
\frac 2n \left( p^{(1)}_n \tr U_1^n  + p^{(2)}_n \tr U_2^n   \right)} d_*U_1 d_*U_2
\ee
This is a different version of the Cauchy matrix model \cite{O-2004-New},\cite{HO-2005} and \cite{Bertola_}.

Instead of (\ref{J1eig}), we obtain
\be\label{J2eig}
K_N(\pb,\pb^*)=\tilde{C}\oint \dots \oint \prod_{i<j}\frac{(u_i-u_j)(v_i-v_j)}{(u_i+u_j)(v_i+v_j)} \prod_{i=1}^N \left(1-c u_i^{-2}v_i^{-2}\right)^{N-1-a}e^{\sum_{n=1,3,5,\dots}
\frac 2n \left( p_n u_i^n  + p^*_n v_i^n   \right)} \frac{du_i}{u_i} \frac{dv_i}{v_i}
\ee
where   $C$ and $\tilde{C}$ provide the condition $K_N(0,0)=1$.

Similarly to the formula (\ref{J1QQ}) in the previous case, we get 
\be\label{J2QQ}
K_N(\pb,\pb^*)=\sum_{\alpha\in DP\atop \ell(\alpha)\le N} Q_{2\alpha}(\pb)Q_{2\alpha}(\pb^*)
\prod_{i=1}^{\ell(\alpha)} \frac{2^{-2\alpha_i}c^{2\alpha_i}(1-c\alpha_i)^{-a}}{\alpha_i!},
\ee
In this case, the formulas (\ref{J1vev}) and (\ref{bosonicJ1}) for the grand partition function are the same; however, in Cauchy's case, we have
\be\label{J2vev}
F(\pb,\pb^*)= -\frac{1}{4\pi^2}\oint\oint
(1-cuv)^{-a}\phi^{(1)}(u,\pb)\phi^{(2)}(v,\pb^*)\frac{du dv}{uv} 
\ee
and
\be\label{bosonicJ2}
F^{\rm Bos}(\hat{\pb},\hat{\pb}^*) =-\frac{1}{4\pi^2} 
\oint\oint (1-cuv)^{-a}V^{(1)}(u,\hat{\pb})V^{(2)}(v,\hat{\pb}^*)\frac{du dv}{uv} 
\ee

Formula (\ref{pf1}) where
\be\label{pf2-pair}
J\left(\pb,\pb(y_i,y_j),\zeta\right)=\sum_N \int_{U_1\times U_1}
e^{\sum_{n>0}
\frac 1n \left( p_n \tr U_1^n\right)}\det\left(1-c U_1^{-2}U_2^{-2}  \right)^{-a}
\det\frac{(1-y_iU_2)(1-y_jU_2)}{(1+y_iU_2)(1+y_jU_2)}  d_*U_1 dU_2^*,
\ee
is still correct.

String equations are the same, namely, (\ref{string_diag}) 
and (\ref{string}),(\ref{M,m}) and (\ref{M,-m}), which are still true, however, now
in (\ref{c}) we have
\be
c(x)=
 \frac{\Gamma(x+m+1)\Gamma(x +a)}{\Gamma(x+1)\Gamma(x +m+a)}\frac{\Gamma(a+m)}{\Gamma(a)} =c_{\rm even}(x)+c_{\rm odd}(x)
 \ee
 For $m=1$, $G\equiv 1$, and $a=1$, we get $c_{\rm even}(x)=1$. Thus, $y_{1,1}=x^3-x=y_{-1,1}$.

\bigskip
\bigskip
\noindent
\small{ {\it Acknowledgements.}
The authors are grateful A. Alexandrov, A.Morozov, A. Mironov for
attracting attention to \cite{Alex2},\cite{MMq}, 
and special thanks to Andrey Mironov for fruitful discussions.
 The work was supported by the Russian Science
Foundation (Grant No.23-41-00049).

\bigskip



\appendix

\section{The Haar measure on $\mathbb{U}_N$ \label{Ensembles}}

The Haar measure on $\mathbb{U}_N$ in its explicit form is written as
\be
d_*U=\frac{1}{(2\pi)^N}\prod_{ 1\le i<k\le N} |e^{\theta_i}-e^{\theta_k}|^2 \prod_{i=1}^N d\theta_i,\quad 
-\pi < \theta_1 < \cdots <\theta_N \le \pi
\ee
where $e^{\theta_1},\dots,e^{\theta_N}$ are eigenvalues of $U\in \mathbb{U}_N$.

\section{Partitions. The Schur and the projective Schur functions}
 
We recall that a non-increasing set of non-negative integers 
$\lambda_1\ge\cdots \ge \lambda_{k}\ge 0$,
we call partition $\lambda=(\lambda_1,\dots,\lambda_{l})$, and $\lambda_i$ are called parts of $\lambda$.
The sum of parts is called the weight $|\lambda|$ of $\lambda$. The number of nonzero parts of $\lambda$
is called the length of $\lambda$, it will be denoted $\ell(\lambda)$. See \cite{Mac} for details.
Partitions will be denoted by Greek letters: $\lambda,\mu,\dots$. The set of all partitions is denoted by
$\Pa$. 
Partitions with distinct parts are called strict partitions, we prefer the
letters $\alpha,\beta$ to denote them. The set of all strict partitions will be denoted by $\DP$.

To define the projective Schur function $Q_\alpha$ where $\alpha\in\DP$, at the first step,
we define the set of functions $\{q_i,i\ge 0\}$ by
\be
e^{\sum_{m>0,{\rm odd}} \frac 2m p_m x^m} =\sum_{m\ge 0} x^m q_m(\pb_{\rm odd})
\ee
where now $\pb_{\rm odd}=(p_1,p_3,\dots)$.
Next, we define the following skew-symmetric matrix:
\be
Q_{ij}(\pb_{\rm odd}) := 
\begin{cases}
q_i(\pb_{\rm odd}) q_j(\pb_{\rm odd}) + 2\sum_{k=1}^j (-1)^kq_{i+k}(\pb_{\rm odd}) q_{j-k}(\pb_{\rm odd}) \quad \text{if } (i,j) \neq (0,0), \\
0  \quad \text{if } (i,j) = (0,0) ,
\end{cases} 
\ee
In particular,
$
Q_{(j,0)}(\pb_{\rm odd}) = -Q_{(0,j)}(\pb_{\rm odd})=q_j(\pb_{\rm odd}) \ \text{ for } j\ge 1.
$
For a strict partition $\alpha=(\alpha_1,\dots,\alpha_{2r})$ where $\alpha_{2r}\ge 0$
the projective Schur function is defined
\be
Q_\alpha(\pb_{\rm odd}):= \Pf\left[Q_{\alpha_i \alpha_j}(\pb_{\rm odd})\right]_{1\le i, j \le 2r},\quad 
Q_{\emptyset} :=1
\label{Q+_pfaff}
\ee
Let $X$ be a matrix.
If we put $p_m=p_m(X)=\ttr\left( X^m - (-X)^m\right)$ which we call odd power sum variables, $m$ odd, we write $Q_\alpha(\pb_{\rm odd}(X))=Q_\alpha(X)$.

\section{Neutral fermions \cite{JM}. Scalar product of the projective Schur functions 
\cite{O-hyp-BKP} \label{scalar}}

For $\alpha=(\alpha_1,\dots,\alpha_k)\in\DP$ introduce
\bea\label{Phi-alpha}
\Phi_\alpha &:=& 2^{\frac k2}\phi_{\alpha_1}\cdots \phi_{\alpha_{k}}\\  
\Phi_{-\alpha} &:=& (-1)^{\sum_{i=1}^k\alpha_i}2^{\frac k2}\phi_{-\alpha_k}\cdots \phi_{-\alpha_{1}}
\eea
We have $\langle 0|\Phi_{-\alpha}\Phi_\beta |0\rangle=2^{\ell(\alpha)}\delta_{\alpha,\beta}$.

The fermionic formula for projective Schur functions was obtained in \cite{You}:
\be\label{Q(x)}
Q_\alpha(\xb)\Delta^*(\xb)=2^{-\frac N2}\langle 0|\phi(-x_1^{-1})\cdots \phi(-x_N^{-1})\Phi_\alpha |0\rangle =2^{-\frac N2}
\langle 0|\Phi_{-\alpha} \phi(x_1)\cdots \phi(x_N) |0\rangle 
\ee
We get
\be
\langle 0|\phi(-v_1^{-1})\cdots \phi(-v_N^{-1})=\Delta^*(v)\sum_{\alpha} 2^{-\frac N2-\ell(\alpha)} Q_\alpha(v) \langle 0|\Phi_{-\alpha} 
\ee
\be
\phi(u_1)\cdots \phi(u_N)|0\rangle =\Delta^*(u)\sum_{\alpha} 2^{-\frac N2-\ell(\alpha)}Q_\alpha(u) \Phi_{\alpha}|0\rangle 
\ee

Following \cite{O-hyp-BKP}
consider $f(x)=\sum_{i\ge 0 }f_i x^i$ and write
\be
2^{-N}\sum_\alpha \Phi_\alpha|0\rangle \langle 0|\Phi_{-\alpha}
\prod_{i=1}^{\ell(\alpha)}  f_{\alpha_i} =
\ee
\be
\frac{1}{(2\pi i)^N}\oint\cdots\oint  \phi(u_1)\cdots \phi(u_N)|0\rangle
\langle 0|\phi(-v_N^{-1})\cdots \phi(-v_1^{-1})
\prod_{i=1}^N\frac{du_idv_i}{u_iv_i}f(u_i^{-1}v_i^{-1})=
\ee
\be
=\frac{1}{(2\pi i)^N}\oint\cdots\oint \Delta^*(u)\Delta^*(v)Q_\alpha(u)Q_\beta(v) \sum_{\alpha,\beta}\Phi_\alpha|0\rangle
\langle 0|\Phi_{-\beta}2^{-N-\ell(\alpha)-\ell(\beta)} \prod_{i=1}^N\frac{du_idv_i}{u_iv_i}f(u_i^{-1}v_i^{-1})
\ee
Then we obtain (\ref{QQproduct}):
\be\label{scal-prod-f}
\frac{1}{(2\pi i)^N}\oint\cdots\oint \Delta^*(u)\Delta^*(v)Q_\alpha(u)Q_\beta(v) 2^{-\ell(\alpha)-\ell(\beta)} \prod_{i=1}^N\frac{du_idv_i}{u_iv_i}f(u_i^{-1}v_i^{-1})=\prod_{i=1}^N f_{\alpha_i}
\ee

\section{Fermionic representation and string equations for the modified $I_N(\pb^{(1)},\pb^{(2)}; f)$ \label{I(f)}}

\be
I_N(\pb^{(1)},\pb^{(2)}
;f)=\langle N,-N|  e^{F(\pb^{(1)},\pb^{(2)})}
|0,0\rangle
\ee
and
\be
F(\pb^{(1)},\pb^{(2)})=-\frac{1}{4\pi^2}\oint\oint f(u^{-1}v^{-1})\psi^{(1)}(u,\pb^{(1)})\psi^{\dag(2)}(v,\pb^{(2)})\frac{du dv}{uv}=
\sum_{j} f_j \psi^{(1)}_j\psi^{\dag(2)}_{-j-1}
\ee
where $f(z)=\sum_j f_j z^j$, $\psi(z)=\sum_j z^j\psi_j$ and $\psi^\dag(z)=\sum_j z^j\psi^\dag_{-j-1}$, and for $j > N$ we have $\langle N|\psi_j|=0$ and $\langle 0|\psi^\dag_{}=0$.
We have
\be
\psi(ze^{\tfrac12 y})\psi^\dag(ze^{-\tfrac12 y})=\sum_{m,n}\Omega_\textsc{f}(m,n) z^{-m} y^n
\ee
where
\be
\Omega_{\textsc{f}}=\sum_{j} (m+\tfrac12)^n\psi_j \psi^\dag_{j+m}
\ee
By $\hat\pb$ we denote the collections $(p_1,p_2,\dots)$ and $(\partial_{p_1},\partial_{p_2},\dots)$ 
The bosonic version is
\be
\Omega_\textsc{b}(m,n,\hat\pb)=
-\res_z \vdots \left(V(z,\hat\pb)z^{-\frac12}\left(z^m D^n z^{\frac12}V(z,-\hat\pb \right) \right) \vdots dz
\ee

Apart of string equations typical for any ``diagonal'' series (\ref{I1-f-series}), string equations which select the prefactor $f$ in (\ref{I1-f-series})
are rather different from the case of two-matrix model with Hermitian matrices.
We have
\be
L=\sum_i x(i)\psi^{(1)}_i\psi^{\dag(1)}_{i+m},
\quad M=\sum_{i} y(i) \psi^{(2)}_i\psi^{\dag(2)}_{i-m},\quad m>0
\ee
and ask
\be\label{cond1}
M|0,0\rangle =0,\quad m>0
\ee
We have
\[
 \sum_i y(i) \psi^{(2)}_i\psi^{\dag(2)}_{i-m}|0,0\rangle=
 y(0) \psi^{(2)}_{0}\psi^{\dag(2)}_{-m}|0,0\rangle +\cdots +
 y_(m-1) \psi^{(2)}_{m-1}\psi^{\dag(2)}_{}{-1}|0,0\rangle=0
\]
thus, $y(i)=0$ for $i=0,\dots,m-1$. Therefore, one can take
\be
y(i)=i(i-1)\dots (i-m+1)
\ee
Then, the condition
\be\label{cond2}
[L+M,\sum_i f_i\psi^{(1)}_i\psi^{\dag(2)}_{-i-1}]=0
\ee
yields
\be
x(i+m)f_{i+m}=y(i)f_i
\ee
Say, for $f_i=\frac{1}{\Gamma(i+1)}$ and $m=1$ one can take $x(i)=i(i-1)$ and $y(i)=i$ (which is
the Virasoro generator denoted $L_{-1}$). 

The bosonized versions are as follows
\be
L^{\textsc{b}}({\hat\tb}^{(1)})
I(\tb^{(1)},\tb^{(2)};f)=
M^{\textsc{b}}({\hat\tb}^{(2)})I(\tb^{(1)},\tb^{(2)};f)
\ee

The explicit examples
will be written down in a more detailed text.

\section{Hirota equations for two-component $BKP$}

We have the following fermionic form for the two-component BKP tau function:
\be
\tau(\pb^{(1)},\pb^{(2)})=\langle 0|
e^{\sum_{i=1,2}\sum_{m>0,{\rm odd}}\frac 2m p_m^{(i)} 
\sum_{j\in Z}\phi^{(i)}_{j-m}\phi^{(i)}_{-j} } h |0\rangle
\ee
where $h$ is of form
\be
h=e^{\sum_{i,j\in \mathbb{Z}}\left(a_{i,j}:\phi^{(1)}\phi^{(1)}:+b_{i,j}:\phi^{(2)}\phi^{(2)}:
+f_{i,j}\phi^{(1)}\phi^{(2)} \right)}
\ee
The fermionic form of the two-component hierarchy is as follows:
\be\label{HirotaF}
\sum_{i=1,2}\sum_{j\in \mathbb{Z}} (-1)^j \phi^{(i)}_j h\otimes \phi^{(i)}_{-j} h - \frac12 h\phi^{(i)}_0\otimes h\phi^{(i)}_0 
\ee
The bilinear Hirota equations are the bosonized version of (\ref{HirotaF}). They can be found in
\cite{KvdLbispec}. In short, one gets
\be
\res_z \left(
\tau\left(\pb^{(1)}+\epsilon^{(1)}-[z],\pb^{(2)}+\epsilon^{(2)}\right)\tau\left(\pb^{(1)}+[z],\pb^{(2)}) \right)\right)
\frac{dz}{z}+
\ee
\[
\res_z \left(
\tau\left(\pb^{(1)}+\epsilon^{(1)},\pb^{(2)}+\epsilon^{(2)}-[z]\right)\tau\left(\pb^{(1)},\pb^{(2)})+[z] \right)\right)
\frac{dz}{z}
-
\tau(\pb^{(1)}+\epsilon^{(1)},\pb^{(2)}+\epsilon^{(2)})\tau(\pb^{(1)},\pb^{(2)})=0
\]
where the sets $\epsilon^{(i)}=\left( \epsilon^{(i)}_1,\epsilon^{(i)}_3,\dots \right)$ are the sets
of arbitrary parameters, and where $[z]$ denotes the set $(2z^{-1},2z^{-3},2z^{-5},\dots)$. The condition of equality to zero for all coefficients of the Taylor series in these parameters gives bilinear differential equations for tau functions called
us the Hirota differential equations.

\end{document}